\documentclass[12pt,a4paper]{article}

\begin{document}
\textwidth=135mm
 \textheight=200mm
\begin{center}
\boldmath
{\bfseries 
All-order $\varepsilon$ expansions of hypergeometric functions of one variable
}
\unboldmath
\vskip 5mm
{\sc Mikhail~Yu.~Kalmykov$^{\dag,\ddag}$} 
and 
{\sc Bernd~A.~Kniehl$^\dag$}
\vskip 5mm
{\small {\it  
{\normalsize $^\dag$ II. Institut f\"ur Theoretische Physik, Universit\"at Hamburg,}\\
{\normalsize Luruper Chaussee 149, 22761 Hamburg, Germany} \\
{\normalsize $^\ddag$ On leave from JINR, 141980 Dubna, Russia.} \\
}} 
\end{center}
\vskip 5mm
\centerline{\bf Abstract}
We briefly sketch a proof concerning the structure of the all-order
$\varepsilon$ expansions of generalized hypergeometric functions with special
sets of parameters. 
\vskip 10mm

{\bf 1.}
Feynman diagrams are the main ingredients for evaluating $S$-matrix elements
within perturbative quantum field theory \cite{BS}.
A powerful technique of dealing with Feynman diagrams is based on their
hypergeometric representations.
However, obtaining exact representations is not enough in practice;
it is also necessary to construct the analytical coefficients of the 
$\varepsilon$ expansions within dimensional regularization in
$d=4-2\varepsilon$ space-time dimensions. 
The first systematic algorithm that is applicable to a large class of
hypergeometric functions with integral values of parameters has recently been
proposed \cite{nested1}, and its generalization to the so-called zero-balance
case has been elaborated \cite{nested2}. 
The resulting expansions are written in terms of Goncharov polylogarithms
\cite{mp}.
This approach allowed one to make the remarkable observation that the
intermediate finite sums, the so-called $Z$ sums, generated by the
$\varepsilon$-expansion procedure form a Hopf algebra.
A similar observation was also made by Kreimer \cite{Kreimer} in the
ultraviolet renormalization procedure in quantum field theory.
However, in physical Feynman diagrams, many other types of sums are generated,
such as multiple (inverse) binomial sums \cite{series,DK2}, 
\begin{equation}
\Sigma^{(k)}_{a1,\cdots,a_p;b_1,\cdots, b_q;c} (z)
= 
\sum_{j=1}^\infty \frac{1}{\left( 2 j \atop j \right)^k} 
\frac{z^j}{j^c} 
S_{a_1}(j\!-\!1) \cdots S_{a_p}(j\!-\!1) 
S_{b_1}(2j\!-\!1) \cdots  S_{b_q}(2j\!-\!1) \; ,
\label{series}
\end{equation}
where $S_a(n)=\sum_{j=1}^n 1/j^a$ is the harmonic sum and $k=\pm 1$.
These sums do not belong to the cases previously studied.
It is, therefore, necessary to develop a new algorithm for the analytical
evaluation of multiple sums of this type and their multivariable 
generalizations. 
A few such approaches were considered (see, for example,
Refs.~\cite{DK2,KWYK,DelDuca}), but a general solution does not yet exist. 
In the following, we briefly describe the approach developed in
Ref.~\cite{KWYK}. 
%
%
%
%
%

\vspace{3mm}

{\bf 2.}
Let us consider the generalized hypergeometric function defined by 
$
{}_{p}F_{p-1}(\vec{A};\vec{B}; z)\!=\!\sum_{j=0}^\infty \frac{\Pi_{i=1}^{p} (A_i)_j}{\Pi_{k=1}^{p-1}(B_k)_j} \frac{z^j}{j!}
$,
where $(A)_j$ is the Pochhammer symbol, $(A)_j = {\Gamma(A+j)}/{\Gamma(A)}$.
Any series of kind~(\ref{series}) can be viewed as a linear combination of 
derivatives of hypergeometric functions with respect to parameters, as 
\begin{equation}
\Sigma^{(k)}_{\vec{a};\vec{b};c}(z)  = 
\sum_{s,\vec{\alpha},\vec{\beta}}
c_s
\left( \partial /  \partial \vec{A}  \right)^{\vec \alpha_s}
\left( \partial /  \partial \vec{B}  \right)^{\vec \beta_s}
\left.
{}_{p+s}F_{p-1+s} \left( \vec{A_s}; \vec{B_s}; z \right)
\right|_{\vec{A_s} = \vec{m_s}; \vec{B_s} = \vec{n_s}}\; ,
\end{equation}
where $\vec{m_s}$ and $\vec{n_s}$ are sets of rational numbers and $c_s$ are
rational functions. 
The problem of analytically evaluating multiple series is reduced to the one
of analytically evaluating the coefficients of the Laurent expansions of
Horn-type hypergeometric functions with respect to their parameters. 

The next step is to apply a differential-reduction algorithm \cite{Takayama}
that allows one to change the value of any parameter of any
hypergeometric function by an arbitrary integer, so that the following
decomposition is valid \cite{Takayama,BKK}: 
\begin{equation}
R_{p+1} {}_{p}F_{p-1}(\vec{A}\!+\!\vec{m};\vec{B}\!+\!\vec{k}; z) = 
\sum_{k=1}^{p} R_k \left( z \frac{d}{dz} \right)^{k-1} {}_{p}F_{p-1}(\vec{A};\vec{B}; z) \;,
\label{decomposition}
\end{equation}
where $\vec{m}$, $\vec{k}$, $\vec{e}_k$, and $\vec{E}_k$ are lists of
integers and $R_k$ are polynomials in the parameters $\vec{A}$, $\vec{B}$, and
$z$.

At this point, it is useful to introduce the polynomials
$P^{(p)}_j(r_1,\cdots,r_p)$ defined as 
\begin{equation}
\prod_{k=1}^p(z+r_k) = 
\sum_{j=0}^p P^{(p)}_{p-j}(r_1,\cdots,r_p) z^j 
\equiv \sum_{j=0}^p P^{(p)}_{p-j}(\vec{r}) z^j 
\equiv \sum_{j=0}^p P^{(p)}_{j}(\vec{r}) z^{p-j} \;,
\label{P}
\end{equation}
so that
\begin{equation}
P^{(p)}_0(\vec{r}) = 1 \;,  
\qquad 
P^{(p)}_j(\vec{r}) =  \sum_{i_1,\cdots,i_r=1}^p \prod_{i_1 < \cdots < i_j }
r_{i_1} \cdots r_{i_j} \;, \qquad j=1, \cdots, p \;.
\end{equation}
For example, 
$P^{(p)}_1(\vec{r}) =  \sum_{j=1}^p r_j$ and
$P^{(p)}_p(\vec{r}) =  \prod_{j=1}^p r_j$.
These polynomials satisfy the following relations: 
\begin{equation}
P^{(p+k)}_{p+k-j}(r_1,\cdots,r_p,q_1,\cdots,q_k) = 
\sum_{n=0}^k P^{(p)}_{p+1-j-n}(r_1,\cdots,r_p) P^{(p)}_{n}(q_1,\cdots,q_k)  \;,
\end{equation}
where  $j = 1, \cdots, p$. 
In particular, we have
$
P^{(p+1)}_{p+1-j}(\vec{r},f) =  P^{(p)}_{p+1-j}(\vec{r}) \!+\! f  P^{(p)}_{p-j}(\vec{r})  \;.
$

Let us consider the $\varepsilon$ expansion of a hypergeometric function with
the following set of parameters:
$
{}_pF_{p-1}\left( \vec{I}\!+\!\vec{a} \varepsilon, A\!+\!c \varepsilon; 
                  \vec{K}\!+\!\vec{b} \varepsilon,B\!+\!\!f \varepsilon;z\right)
$,
where $\vec{I}$ and $\vec{K}$ are integers and $A$, $B$, $\vec{a}$, $\vec{b}$,
$c$, and $f$ are arbitrary rational numbers. 
In accordance with Eq.~(\ref{decomposition}), this function can be written as
a linear combination of $p-1$ differential operators acting on the
hypergeometric function $\omega(z)$ with the following set of parameters:
$
\omega(z) = 
{}_pF_{p-1}\left( \vec{a} \varepsilon,A\!+\!c \varepsilon; \vec{1}\!+\!\vec{b} \varepsilon,B\!+\!\!f \varepsilon;z\right)
$.
Starting from the differential equation for $\omega(z)$,
\begin{equation}
\left[ 
z \left(\theta\!+\!A\!+\!c\varepsilon\right)  \Pi_{j=1}^{p-1}(\theta\!+\!a_j\varepsilon) 
\!-\! 
\theta 
\left(\theta\!+\! B\!-\!1 \!+\!f\varepsilon\right)
\Pi_{k=1}^{p-2} (\theta\!+\!b_k\varepsilon)
\right] \omega(z) = 0 \;,
\label{de}
\end{equation}
and writing its $\varepsilon$ expansion as
$
\omega(z) = 1 + \sum_{j=1}^\infty w_k(z) \varepsilon^k 
$, 
we obtain the following system of differential equations for $\{w_m(z)\}$:
\begin{eqnarray}
\lefteqn{
\left[ 
(1\!-\!z) \frac{d}{dz} \!+\! \frac{B\!-\!1}{z} \!-\! A 
\right] 
\theta^{p-1} w_m(z) 
= 
\left[ 
P^{(p)}_{1}(\vec{a},c)
\!-\! 
\frac{1}{z} 
P^{(p-1)}_{1}(\vec{b},f)
\right]
\theta^{p-1} w_{m-1}(z)}
\nonumber \\
&&{} +
\sum_{j=2}^{p-1}
\left[ 
P^{(p)}_{j}(\vec{a},c)
- 
\frac{1}{z} 
P^{(p-1)}_{j}(\vec{b},f)
\right]
\theta^{p-j} w_{m-j}(z) 
+ A P^{(p-1)}_{p-1}(\vec{a}) w_{m-p+1}(z)
\nonumber \\
&&{} + 
\sum_{k=1}^{p-2}
\left[ 
A P^{(p-1)}_{k}(\vec{a})
\!-\! 
\frac{(B-1)}{z} 
P^{(p-2)}_{k}(\vec{b})
\right]
\theta^{p-1-k} w_{m-k}(z)
\!+\! P^{(p)}_{p}(\vec{a},c) w_{m-p}(z) \;,
\nonumber\\
\label{eq1}
\end{eqnarray}
where $\theta=zd/dz$. 
The first non-vanishing term corresponds to $m=p$ if $A = 0$ and $m=p-1$
otherwise. 
In both cases, Eq.~(\ref{eq1}) reduces to 
\begin{equation}
\left[ 
(1\!-\!z) \frac{d}{dz} \!+\! \frac{B\!-\!1}{z} \!-\! A 
\right] 
\theta^{p-1} w_{p-1+\delta_{A,0}}(z)  = 
(A \!+\! c \delta_{A,0} ) P^{(p-1)}_{p-1}(\vec{a}) \;,
\label{zero}
\end{equation}
where $\delta_{A,0} $ is equal to $1$ if $A=0$ and zero otherwise.
To simplify Eq.~(\ref{zero}), let us redefine the higher derivatives of
$\omega(z)$ as
$\theta^{p-1} w_k(z) \to h(z) \theta^{p-1} \phi_k(z)$, 
where $\phi_k(z)$ is a new function and 
\begin{equation}
h(z) = (-1)^A z^{1-B}(z-1)^{B-A-1}\;,
\label{h}
\end{equation}
with $A$ and $B$ being arbitrary rational numbers. 
Then, Eq.~(\ref{zero}) becomes
\begin{eqnarray}
(-1)^{A-1} z^{-B}(z-1)^{B-A} \theta^{p} \phi_{p-1+\delta_{A,0}}(z) = 
( A \!+\! c \delta_{A,0}) P^{(p-1)}_{p-1}(\vec{a}) \;.
\label{zero:1}
\end{eqnarray}
The solution of Eq.~(\ref{zero:1}) can be written as a multiply iterated
integral,
\begin{equation}
\phi_{p-1+\delta_{A,0}(z)}^{(p-1)} \sim 
\int_0^z \frac{dt_1}{t_1} \int_0^{t_1} \frac{dt_2}{t_2} 
\cdots 
\int_0^{t_{p-1}} \frac{dt_p}{t_p} \frac{t_p^B}{(t_p-1)^{B-A}} \;,
\label{sol1}
\end{equation}
where the constant part is omitted for simplicity.
This solution can be written in terms of hyperlogarithms
defined as iterative integrals over rational one-forms, 
\begin{equation}
I_k(z; a_k, a_{k-1},\ldots , a_1) =  
\int_0^{z} \frac{dt}{t\!-\!a_k}
I_{k-1}(t;a_{k-1},\ldots , a_1)
\;,
\label{I}
\end{equation}
where $z$ is the argument, $\{a_i\}$ is the set of parameters, and $k$ is the
weight of the hyperlogarithm.
In this way, the solution in the form of Eq.~(\ref{sol1}) may be expressed in
terms of hyperlogarithms if a parametrization $z \to \xi(z)$ exists such that  
following two conditions are fulfilled:
\begin{equation}
\frac{dz}{(1-z) h(z) }= Q(\xi) d \xi \;,
\qquad 
\frac{dz}{z } = R(\xi) d \xi \;,
\label{reparametrization}
\end{equation}
where $Q(\xi)$ and $R(\xi)$ are rational functions of $\xi$. 
Using the parametrization $A=r/q$ and $B=1-p/q$, where $p$, $r$, and $q$
are integers, the three most important cases are: 
(i) $A=0$, $B=1-p/q$, $\{[(z/(z-1)]^{p/q}\}$;
(ii) $A=r/q$, $B=1$, $\{(1-z)^{-r/q}\}$;
(iii) $B-A = k$, $\{(1-z)^{k-1} z^{p/q}\}$,
where $k$ is integer and the function $h(z)$ is written out in braces.
The new variables $\xi$ for these cases may be chosen as \cite{nested2} 
(i) $\xi = [z/(z-1)]^{1/q}$;
(ii) $\xi=(1-z)^{1/q}$;
(iii) $\xi= z^{1/q}$.
We point out that another parametrization exists for $q=2$ \cite{series,DK2}.

\noindent 
{\bf Remark A.}
It is easy to show that Eq.~(\ref{reparametrization}) is equivalent to the
statement that the hypergeometric function 
$
z {}_pF_{p-1}\left(1\!+\!A\!,\vec{1}_{p-1}; 1\!+\!B,\vec{2}_{p-2} ;z\right)
$
is expressible in terms of rational functions times hyperlogarithms. 

In order to analyze the structure of the highest coefficients of the
$\varepsilon$ expansions, let us consider the original function $\omega(z)$
and its first $p\!-\!1$ derivatives as independent functions,
$f^{(k)} = (\omega,  \theta \omega, \cdots,  \theta^{p-1} \omega )$,
$k = 0,\cdots, p-1$.
Taking into account that each of the functions $ f^{(k)}$ has a $\varepsilon$
expansion of the form 
$f^{(k)}(z) = \sum_{j=0}^\infty f_j^{(k)}(z) \varepsilon^j$
with the boundary conditions 
$f_0^{(0)}(z)=1$ and  $f_j^{(k)}(0)=0$, $j \geq 1$, $k=1,\cdots, p-1$
and redefining
$\theta^{p-1} \omega_k(z) = h(z) \phi_j^{(p-1)}(z)$,
we convert Eq.~(\ref{eq1}) into a system of first-order differential
equations, 
\begin{eqnarray}
\lefteqn{h(z) (1\!-\!z) \frac{d}{dz} \phi_m^{(p-1)}(z)
= 
h(z) 
\left[ 
P^{(p)}_{1}(\vec{a},c)
\!-\! 
\frac{1}{z} 
P^{(p-1)}_{1}(\vec{b},f)
\right]
\phi_{m-1}^{(p-1)}(z)}
\nonumber \\ 
&&{}+
\sum_{j=2}^{p-1}
\left[ 
P^{(p)}_{j}(\vec{a},c)
- 
\frac{1}{z} 
P^{(p-1)}_{j}(\vec{b},f)
\right]
f_{m-j}^{(p-j)}(z)
+ A P^{(p-1)}_{p-1}(\vec{a}) w_{m-p+1}(z)
\nonumber\\
&&{}+ 
\sum_{k=1}^{p-2}
\left[ 
A P^{(p-1)}_{k}(\vec{a})
\!-\! 
\frac{(B-1)}{z} 
P^{(p-2)}_{k}(\vec{b})
\right]
 f_{m-k}^{(p-1-k)}(z)
\!+\! P^{(p)}_{p}(\vec{a},c) w_{m-p}(z) \;,
\nonumber\\
&&\theta  f^{(p-2)}_{m}(z) = h \phi_{m}^{(p-1)}(z) \;,
\nonumber\\
&&\theta f_{m}^{(j-1)}(z)  =  f_{m}^{(j)}(z) \;,
\qquad j = 1, \cdots, p-2 \;.
\end{eqnarray}
The solution of this system can again be presented as an iterated integral
over a rational one form, if two additional conditions are satisfied: 
\begin{equation}
\frac{dz}{z}\, \frac{1}{h(z) } =  P_1(\xi) d \xi \;,
\qquad 
\frac{dz}{z}h(z) = P_2(\xi) d \xi \;,
\label{reparametrization2}
\end{equation}
where $P_1$ and $P_2$ are rational functions. 
As a consequence of the universality of hyperlogarithms, any iterated integral
over a rational function may be expressed again in terms of hyperlogarithms. 
It is easy to show that the two equations in Eq.~(\ref{reparametrization2})
are not functionally independent.
In fact, using the second equality in Eq.~(\ref{reparametrization}), we obtain 
$R^2(\xi) = P_1(\xi) P_2(\xi)$ 
and
$h(z) = R(\xi)/P_1(\xi) =  P_2(\xi)/R(\xi)$.

\noindent
{\bf Remark B.}
In Ref.~\cite{nested2}, the zero-balance case was analyzed via the algebra of
nested sums, and it was proven that the coefficients of the $\varepsilon$
expansion are expressible in terms of hyperlogarithms of $q$-roots of unity
with argument $z^{1/q}$.
Also, the proposition was made that any hypergeometric function with one
unbalanced rational parameter is again expressible in terms of hyperlogarithms
of $q$-roots of unity with arguments $[z/(z-1)]^{1/q}$ or $(1-z)^{1/q}$.
But this statement is in contradiction with the results of
Ref.~\cite{DK2,LL04}, which were confirmed later in Ref.~\cite{HM}. 

\noindent 
{\bf Remark C.}
In Ref.~\cite{HM}, an ansatz for the coefficients of the $\varepsilon$
expansions of ${}_3F_2$ hypergeometric functions was presented, and it was
shown that the first few terms are compatible with the differential equations 
for the hypergeometric functions. 
However, the proof of validity of this ansatz for an arbitrary order of
$\varepsilon$ was not delivered.

\noindent 
{\bf Conclusions.}
The analytical structure of the coefficients of the all-order $\varepsilon$
expansion of the hypergeometric function 
$
{}_pF_{p-1}\left( \vec{I}\!+\!\vec{a} \varepsilon, A\!+\!c \varepsilon; 
                  \vec{K}\!+\!\vec{b} \varepsilon,B\!+\!\!f \varepsilon;z\right)$,
where $\vec{I}$ and $\vec{K}$ are integers and $A$, $B$, $\vec{a}$, $\vec{b}$,
$c$, and $f$ are arbitrary rational numbers, was analyzed.
It was shown that, under the conditions of Eq.~(\ref{reparametrization}) and
one of those of Eq.~(\ref{reparametrization2}), the coefficients are
expressible in terms of hyperlogarithms with arguments and parameters defined
through three polynomials $R$, $Q$, and $P_1$. 

\noindent 
{\bf Acknowledgments.}
One of us (M.Yu.K) it grateful to the organizers of the 2009 International
Bogoliubov Conference on Problems of Theoretical and Mathematical Physics and
to Dmitry Kazakov for the inviation and partial financial support.
This work was supported in part by BMBF Grant No.\ 05H09GUE,
DFG Grant No.\ KN~365/3--2, and HGF Grant No.\ HA~101.


\begin{thebibliography}{99}

\bibitem{BS}
\textit{Bogoliubov N.N. and Shirkov D.V.} //
{Introduction to the Theory of Quantized Fields}
(Wiley \& Sons, New York, 1980).

\bibitem{nested1}
\textit{Moch S., Uwer P. and Weinzierl S.} //
{J. Math. Phys. 2002. V.43. P.3363.}

\bibitem{nested2}
\textit{Weinzierl S.} //
{J. Math. Phys. 2004. V.45. P.2656.}

\bibitem{mp}
\textit{Goncharov A.B.} //
{Math. Res. Lett. 1997. V.4. P.617;} \\
\textit{Remiddi E. and Vermaseren J.A.M.} //
{Int. J. Mod. Phys. A. 2000. V.15. P.725;} \\
\textit{Borwein J.M. et al.} //
{Trans. Amer. Math. Soc. 2001. V.353. P.907;} \\
\textit{Vollinga J. and Weinzierl S.} //
{Comput. Phys. Commun. 2005. V.167. p.177.}

\bibitem{Kreimer}
\textit{Kreimer D.} //
{Adv. Theor. Math. Phys. 1998. V.2. P.303.}

\bibitem{series}
\textit{Broadhurst D.J.} //
{Eur. Phys. J. C. 1999. V.8. P.311;} \\
\textit{Davydychev A.I. and Kalmykov M.Yu.} //
{Nucl. Phys. B. 2001. V.605. P.266;} \\
\textit{Jegerlehner F., Kalmykov M.Yu. and Veretin O.} //
{Nucl. Phys. B. 2003. V.658. P.49.} 



\bibitem{DK2}
\textit{Davydychev A.I. and Kalmykov M.Yu.} //
{Nucl. Phys. B. 2004. V.699. P.3.}  


\bibitem{KWYK}
\textit{Kalmykov M.Yu., Ward B.F.L. and Yost S.} // 
{JHEP. 2007. 02.040;} 
{JHEP. 2007. 10.048;}
{JHEP. 2007. 11.009;} \\  
\textit{Kalmykov M.Yu. and Kniehl B.A.} // 
{Nucl. Phys. B. 2009. V.809. P.365;} \\
\textit{Kalmykov M.Yu. et. al.}  // arXiv:0810.3238.

\bibitem{DelDuca}
\textit{Del Duca V. et al.} //
{JHEP. 2010. 01.042.}

\bibitem{Takayama}
\textit{Takayama N.} //
Japan J. Appl. Math. 1989. V.6. P.147.

\bibitem{BKK}
\textit{Kalmykov M.Yu.}//
{JHEP. 2006. 04.056;} \\ 
\textit{Bytev V.V., Kalmykov M.Yu. and Kniehl, B.A.} //
arXiv:0904.0214.

\bibitem{LL04}
\textit{Kalmykov M.Yu.} //
{Nucl. Phys. Proc. Suppl. 2004. V.135. P.280.}

\bibitem{HM}
\textit{Huber T. and Maitre D.} //
{Comput. Phys. Commun. 2008. V.178. P.755.}

\end{thebibliography}
\end{document}